\documentclass[twocolumn,prd,groupedaddress,nofootinbib]{revtex4}
\usepackage{amssymb}
\usepackage{latexsym}

\input{tcilatex}

\begin{document}

\title{Analogue of superradiance effect in acoustic black hole in the
presence of disclination}
\author{Geusa de A. Marques}
\date{\today }

\begin{abstract}
In this paper we invstigate the possibility of the acoustic analogue of a
phenomenon like superradiance, that is, the amplification of a sound wave by
reflection from the ergo-region of a rotating acoustic black hole in the
fluid "draining bathtub" model in the presence of a desclination be
amplified or reduced in agreement with the value of the deficit angle.

\smallskip
\end{abstract}

\affiliation{Unidade Acad\^{e}mica de F\'{\i}sica, Universidade Federal de Campina Grande%
\\
58109-790, Caixa Postal 10071, Campina Grande, Pb, Brazil.\\
\medskip \\
Electronic address: gmarques@df.ufcg.edu.br\\
\smallskip }
\maketitle

\section{Introduction}

Acoustic analogue of a black hole has been a lot studied in the literature
as a concrete laboratory model for probe several aspects of curved space
quantum field theory.

In 1981, Unruh\cite{1,2} showed that if a fluid is barotropic and inviscid,
and the flow of the fluid is irrotational, the equation of motion that
fluctuation of the velocity potential of acoustic disturbance obeys, is
identical to that of a minimally coupled massless scalar field propagating
in an effective curved spacetime Loretizian geometry, which can simulate an
artificial black hole\cite{1,2}.

This paper is organized as follows. In section 2, we obtain the effective
acoustic geometry. In section 3, we will show the Klein-Gordon equation in
the sonic black hole scenario. In section 4, we describe the acoustic black
hole in the presence of a disclination and the amplification sound wave.
Section 5 is devoted to present our conclusions.

\section{Effective acoustic geometry}

In the absence of chemical reactions, the number of particles entering and
leaving a collision in a fluid will be conserved. For non relativistic
process, the total mass of the particles involved in the collision process
will also be conserved. As a result, if we consider a volume element of the
fluid, $dV(t)$ (with a given set of fluid particles), which moves with the
fluid, the amount of mass inside this volume element must remain constant.
Let be $\rho =\rho (\vec{r},t)$ the mass density (mass per unit volume) and
let $M$ denote the total mass in the volume, $V(t)$. Then%
\begin{equation}
\frac{dM}{dt}=\frac{d}{dt}\dint\limits_{V(t)}\rho
dV=\dint\limits_{V(t)}\left( \frac{d\rho }{dt}+\rho \vec{\nabla}_{r}\cdot 
\vec{v}\right) dV=0,  \label{1}
\end{equation}%
where $\vec{v}=\vec{v}(\vec{r},t)$ is the average velocity of the fluid at
point $\vec{r}$ and time $t$. Since the volume \ element, $dV(t)$, is
arbitrary, the integrand must be zero and we find%
\begin{equation}
\frac{d\rho }{dt}+\rho \vec{\nabla}_{r}\cdot \vec{v}=0.  \label{2}
\end{equation}%
If we that the convective derivative is given by $d/dt=\partial /\partial t+$
$\vec{v}\cdot \vec{\nabla}_{r}$, then we can also write%
\begin{equation}
\frac{\partial \rho }{\partial t}+\vec{\nabla}_{r}\cdot \left( \rho \vec{v}%
\right) =0,  \label{3}
\end{equation}%
where $\rho \vec{v}$ is the mass flux. The eq. (\ref{3}) is a continuity
equation and it is a direct consequence of the conservation of mass in the
fluid. The Euler equation is given by%
\begin{equation}
\rho \frac{d\vec{v}}{dt}\equiv \rho \lbrack \frac{\partial \vec{v}}{\partial
t}+(\vec{v}\cdot \vec{\nabla})\vec{v}]=\vec{f}+\vec{F},  \label{4}
\end{equation}%
where $f$ is a force per unit volume acting on the walls of the volume
element and $F$ is an external force per unit volume which couples to the
particles inside the volume element (it could be an electric or magnetic
field for example).

We consider a fluid been inviscid (zero viscosity), with the only forces
present being those due to pressure $p$, i.e., $\vec{f}=-\vec{\nabla}p$. In
this case, $\vec{F}$ equal to zero. In the sense, we consider that the fluid
is locally irrotational (free vortex), that is, $\vec{v}=-\vec{\nabla}\phi ,$
and the fluid is barotropic, i.e., the density $\rho $ is a function of
pressure $p$ only. In this case, we can define the enthalpy $h$ as%
\begin{equation}
h(p)=\dint\limits_{0}^{p}\frac{dp^{\prime }}{\rho (p^{\prime })}  \label{5}
\end{equation}%
or%
\begin{equation}
\vec{\nabla}h=\frac{\vec{\nabla}p}{\rho }.  \label{6}
\end{equation}%
The equation (\ref{4}) now reduce to 
\begin{equation}
-\frac{\partial \phi }{\partial t}+h+\frac{1}{2}\left( \vec{\nabla}\phi
\right) ^{2}=0.  \label{7}
\end{equation}

We will follow to study sound wave, the usual procedure and linearize the
continuity and Euler 's equations around some background flow, by setting $%
\rho =\rho _{0}+\epsilon \rho _{1},$ $p=p_{0}+\epsilon p_{1},$ $\phi =\phi
_{0}+\epsilon \phi _{1}$, and discarding all terms of order $\epsilon ^{2}$
or higher.

Then, the continuity equation leads to%
\begin{equation}
\frac{\partial \rho _{0}}{\partial t}+\vec{\nabla}\cdot (\rho _{0}\vec{v}%
_{0})=0.  \label{8}
\end{equation}%
and%
\begin{equation}
\frac{\partial \rho _{1}}{\partial t}+\vec{\nabla}\cdot (\rho _{1}\vec{v}%
_{0}+\rho _{0}\vec{v})=0,  \label{9}
\end{equation}

Expanding $h(p)$ as $h(p_{0}+\epsilon p_{1})\simeq h(p_{0})+\epsilon \frac{dh%
}{dp}|_{p=p_{0}}=h_{0}+\epsilon \frac{p_{1}}{\rho _{0}}=h_{0}+\epsilon
h_{1}, $ the eq. (\ref{4}) becomes%
\begin{equation}
-\frac{\partial \phi _{0}}{\partial t}+h_{0}+\frac{1}{2}(\vec{\nabla}\phi
_{0})^{2}=0,  \label{10}
\end{equation}%
and%
\[
-\frac{\partial \phi _{1}}{\partial t}+\frac{p_{1}}{\rho _{0}}-\vec{v}%
_{0}\cdot \vec{\nabla}\phi _{1}=0, 
\]%
that is%
\begin{equation}
p_{1}=\rho _{0}\left( \frac{\partial \phi _{1}}{\partial t}+\vec{v}_{0}\cdot 
\vec{\nabla}\phi _{1}\right) .  \label{11}
\end{equation}%
Then, since the fluid is barotropic we have%
\begin{equation}
\rho _{1}=\frac{\partial \rho }{\partial p}p_{1}.  \label{12}
\end{equation}%
Substituting eq. (\ref{11}) into (\ref{12}) we get%
\begin{equation}
\rho _{1}=\frac{\partial \rho }{\partial p}\rho _{0}\left( \frac{\partial
\phi _{1}}{\partial t}+\vec{v}_{0}\cdot \vec{\nabla}\phi _{1}\right) .
\label{13}
\end{equation}%
Now, substituting eq. (\ref{13}) into eq. (\ref{9}) we obtain

\begin{widetext}
\begin{eqnarray}
\label{14}
-\frac{\partial }{\partial t}\left[ \frac{\partial \rho }{\partial p}\rho
_{0}\left( \frac{\partial \phi _{1}}{\partial t}+\vec{v}_{0}\cdot \vec{\nabla%
}\phi _{1}\right) \right] +\vec{\nabla}\cdot \left[ \rho _{0}\vec{\nabla}%
\phi _{1}-\frac{\partial \rho }{\partial p}\rho _{0}\vec{v}_{0}\left( \frac{%
\partial \phi _{1}}{\partial t}+\vec{v}_{0}\cdot \vec{\nabla}\phi
_{1}\right) \right] =0.
\end{eqnarray}

\end{widetext}The eq. (\ref{14}) describes the propagation of the linearized
scalar potential $\phi _{1},$ if $\phi _{1}$ is determined, eq. (\ref{11})
determines $\rho _{1}$. Thus, this wave equation completely determines the
propagation of acoustic disturbances, where the local speed of sound is
defined by%
\begin{equation}
c^{-2}\equiv \frac{\partial \rho }{\partial p}.  \label{15}
\end{equation}

Thus, it can now be shown that the eq. (\ref{14}) can also be obtained from
the usual curved space Klein Gordon equation\cite{wisser}%
\begin{equation}
\frac{1}{\sqrt{-g}}\partial _{\mu }\left( \sqrt{-g}g^{\mu \nu }\partial
_{\nu }\right) \phi =0,  \label{16}
\end{equation}%
where $g_{\mu \nu }$ is a metric tensor (with Lorentizian signature), not of
spacetime itself. but an acoustic $\ ^{\prime \prime }$analog spacetime$%
^{\prime \prime }$.

\section{Klein-Gordon equation in the sonic black hole scenario - Draining
bathtub flow model}

In this model, the velocity potential, in polar coordinates is given by\cite%
{wisser}%
\begin{equation}
\phi (r,\theta )=A\log r+B\theta ,  \label{17}
\end{equation}%
where $A$ and $B$ are real constants and $\phi $ present a sink in the
origin. This leads to the velocity profile%
\begin{equation}
\vec{v}=\frac{A}{r}\hat{r}+\frac{B}{r}\hat{\theta},  \label{18}
\end{equation}%
then, the metric in the exterior region, i.e., outside of core at $r=0$,
turns out 
\begin{widetext}
\begin{eqnarray}
ds^{2}=-\left( c^{2}-\frac{A^{2}+B^{2}}{r^{2}}\right) dt^{2}-\frac{2A}{r}%
drdt-2Bd\theta dt+dr^{2}+r^{2}d\theta ^{2}+dz^{2},  \label{19}
\end{eqnarray}
\end{widetext}where $c$ is the velocity of sound.

Defining\cite{sou}%
\[
dt\rightarrow dt+\frac{\left\vert A\right\vert r}{\left(
r^{2}c^{2}-A^{2}\right) }dr\text{ ; \ \ }d\theta \rightarrow d\theta +\frac{%
B\left\vert A\right\vert r}{r\left( r^{2}c^{2}-A^{2}\right) }dr\text{\ \ \ \
\ \ } 
\]%
we obtain, after a rescaling of the time coordinate by $c$ 
\begin{widetext}
\begin{eqnarray}
ds^{2}=-\left( 1-\frac{A^{2}+B^{2}}{c^{2}r^{2}}\right) dt^{2}-\left( 1-\frac{%
A^{2}}{c^{2}r^{2}}\right) ^{-1}dr^{2}-2\frac{B}{c}d\theta dt+r^{2}d\theta
^{2}+dz^{2},  \label{20}
\end{eqnarray}
\end{widetext}where , the radius of the ergosphere is given by the vanishing
of $g_{00}$, i.e., $r_{e}=\left( A^{2}+B^{2}\right) ^{1/2}/c$, and it has a
singularity at $r_{h}=\left\vert A\right\vert /c$, which signifies the event
horizon. We observe on eq. (\ref{18}) that for $A>0$ we are dealing with a
past event horizon, i.e., acoustic white hole and for $A<0$ we dealing with
a future acoustic horizon, i.e., acoustic black hole.

\section{Acoustic black hole in the presence of a disclination and
amplification sound wave}

In a recent paper\cite{F} we have discussed the phenomenon of sound
amplification in the acoustic black hole analogue.

Thus, in this paper we propose the analyze the influence of an acoustic
black hole analogue in the presence of a disclination in the sound wave
amplification. In the geometric approach, the medium with a disclination has
the line element given by

\begin{widetext}
\begin{eqnarray}
ds^{2}=-\left( 1-\frac{A^{2}+B^{2}}{c^{2}r^{2}}\right) dt^{2}-\left( 1-\frac{%
A^{2}}{c^{2}r^{2}}\right) ^{-1}dr^{2}-2\frac{B}{c}\alpha d\theta dt+r^{2}\alpha^2 d\theta
^{2}+dz^{2},  \label{21}
\end{eqnarray}
\end{widetext}in cylindrical coordinates. This metric is equivalent to the
boundary condition with periodicity of $2\pi \alpha $ instead of $2\pi $
around the $z-$axis. In the Volterra process\cite{Kle} of disclination
creation, this corresponds to remove $(0<\alpha \leq 1)$ or insert $(2\pi
>\alpha \geq 1)$ a wedge of material of dihedral angle $\lambda =2\pi
(\alpha -1)$\cite{G}.

But, for the velocity potential given by eq. (\ref{18}), the analogue black
hole metric is basically a $(2+1)$ dimensional flow with a sink at the
origin. The metric given by (\ref{21}) reduce to 
\begin{widetext}
\begin{eqnarray}
ds^{2}=-\left( 1-\frac{A^{2}+B^{2}}{c^{2}r^{2}}\right) dt^{2}-\left( 1-\frac{%
A^{2}}{c^{2}r^{2}}\right) ^{-1}dr^{2}-2\frac{B}{c}\alpha d\theta dt+r^{2}\alpha^2 d\theta
^{2}. \label{22}
\end{eqnarray}
\end{widetext}

Now, we write the Klein-Gordon equation (\ref{16}) in the background metric (%
\ref{22}) and we can separate variables by the substitution%
\[
\phi (t,r,\theta )=\exp i(\omega t-m\theta )R(r), 
\]%
where $m$ is an integer, we assume that $\omega >0$, then, the radial
function satisfies the equation given by

\begin{widetext}
\begin{eqnarray}
\frac{1}{r}\left( 1-\frac{A^{2}}{c^{2}r^{2}}\right) \frac{d}{dr}\left[
r\left( 1-\frac{A^{2}}{c^{2}r^{2}}\right) \frac{d}{dr}\right] R(r)+\left[
\omega ^{2}-\frac{2Bm\omega }{\alpha cr^{2}}-\frac{m^{2}}{\alpha ^{2}r^{2}}%
\left( 1-\frac{A^{2}+B^{2}}{c^{2}r^{2}}\right) \right] R(r)=0.
\label{24}
\end{eqnarray}
\end{widetext}

Introducing the tortoise coordinate $r^{\ast }$ such that%
\begin{equation}
\frac{d}{dr^{\ast }}=\left( 1-\frac{A^{2}}{r^{2}c^{2}}\right) \frac{d}{dr}
\label{25}
\end{equation}%
which implies that%
\begin{equation}
r^{\ast }=r+\frac{|A|}{2c}\log \left\vert \frac{r-\frac{|A|}{c}}{r+\frac{|A|%
}{c}}\right\vert .  \label{26}
\end{equation}%
Observe that the horizon $r=\frac{|A|}{c}$ maps to $r^{\ast }\rightarrow
-\infty $ and while $r\rightarrow \infty $ corresponds to $r^{\ast
}\rightarrow +\infty $. Now, introducing a new radial function $g(r^{\ast
})\equiv r^{1/2}R(r),$ we obtain the equation%
\begin{equation}
\frac{d^{2}g(r^{\ast })}{dr^{\ast 2}}+\left[ q(r)-\frac{1}{2r^{2}}\left( 
\frac{dr}{dr^{\ast }}\right) ^{2}-\left( \frac{A^{2}}{r^{4}c^{2}}-\frac{3}{%
4r^{2}}\right) \frac{dr}{dr^{\ast }}\right] g(r^{\ast })=0,  \label{27}
\end{equation}%
where%
\begin{equation}
q(r)=\frac{A^{2}m^{2}+Bm^{2}-c^{2}m^{2}r^{2}-2B\alpha mr^{2}\omega +\alpha
^{2}r^{4}\omega ^{2}}{c^{2}\alpha ^{2}r^{4}}.  \label{28}
\end{equation}%
Now, analyzing eq. (\ref{27}) when $r\rightarrow \infty $, we obtain%
\begin{equation}
\frac{d^{2}g(r^{\ast })}{dr^{\ast 2}}+\omega ^{2}g(r^{\ast })=0,  \label{29}
\end{equation}%
whose solution is given by%
\begin{equation}
g(r^{\ast })=\exp \left( i\omega r^{\ast }\right) +\mathcal{R}\exp (-i\omega
r^{\ast }).  \label{30}
\end{equation}%
The first term of eq. (\ref{30}) corresponds to an ingoing wave and the
second term corresponds to de reflected wave, where $\mathcal{R}$ is the
reflection coefficient in the sense of potential scattering. Now using this
solution of the differential equation together with it complex conjugate, we
calculate the Wronskian of the solutions (\ref{30}) given by%
\begin{equation}
\mathcal{W}(+\infty )=-2i\omega \left( 1-|\mathcal{R}|^{2}\right) .
\label{31}
\end{equation}

Thus, we considering the solution near the horizon, is that, $r^{\ast
}\rightarrow -\infty ,$ the eq. (\ref{27}) becomes%
\begin{equation}
\frac{d^{2}g(r^{\ast })}{dr^{\ast 2}}+\left( \omega -m\Omega _{H,\alpha
}\right) ^{2}g(r^{\ast })=0  \label{32}
\end{equation}%
where $\Omega _{H,\alpha }\equiv \frac{Bc}{\alpha A^{2}}$ is the angular
velocity of the acoustic black hole in the presence of a disclination. Near
the horizon, we suppose that just the solution identified by ingoing wave is
physical, is that%
\begin{equation}
g(r^{\ast })=\mathcal{T}\exp [i\left( \omega -m\Omega _{H,\alpha }\right)
r^{\ast }],  \label{33}
\end{equation}%
where $\mathcal{T}$ is the transmission coefficient. Once again, we
calculate the Wronskian of the solutions (\ref{33})

\begin{equation}
\mathcal{W}(-\infty )=-2i\left( \omega -m\Omega _{H,\alpha }\right)
\left\vert \mathcal{T}\right\vert ^{2}.  \label{34}
\end{equation}%
Thus, remind that two linearly independent solutions of the same
differential equation must lead to a constant Wronskian, so of eqs. (\ref{31}%
) and (\ref{34}) we obtain%
\begin{equation}
\left\vert \mathcal{R}\right\vert ^{2}=1-\left( 1-\frac{m}{\omega }\Omega
_{H,\alpha }\right) \left\vert \mathcal{T}\right\vert ^{2}.  \label{35}
\end{equation}%
We can observe in eq. (\ref{35}) that, for frequencies in the range $%
0<\omega <$ $m\Omega _{H,\alpha }$, the reflection coefficient has a
magnitude larger than unity whose imply the amplification relation of the
ingoing sound wave near horizon regions. This imply that the ingoing wave
removes mass (energy) of the acoustic black hole\cite{F}. As $\Omega
_{H,\alpha }$ depends on the disclination, then, the same affects the
quantity of removed energy of the hole. When $0<\alpha \leq 1$ whose
corresponds to remove a wedge of material it is possible to accentuate the
quantity of retired energy of the acoustic black hole, in other words,
larger amplification of the ingoing sound wave and when $2\pi >\alpha \geq 1$
whose corresponds to insert a wedge of material represent to the possibility
to attenuate the quantity of removed energy of the acoustic black hole.

\section{Conclusions}

In this paper we shown that the presence of the disclination modify the
quantity of removed energy of the acoustic black hole and that, it is
possible to accentuate or to attenuate the amplification of the removed
energy of the acoustic black hole and still exists the possibility to cancel
the superradiance effect to $\alpha $ equal to $m\Omega /\omega $ where $%
\Omega $ $\equiv \frac{Bc}{A^{2}}$ is the angular velocity of the acoustic
black hole in the absence of the disclination, in this case, the reflection
coefficient is equal to unity. Those aspects perhaps can be proven in future
experimental realizations.

\vskip2.0 cm

{\large \textbf{Acknowledgments:}} I would like thank to MCT/CNPq and
FAPESQ-PB for the partial financial support.

\end{document}